\documentclass[reprint,
superscriptaddress,
 amsmath,amssymb,
 aps,
prb,
]{revtex4-2}

\usepackage{graphicx}
\usepackage{dcolumn}
\usepackage{bm}
\usepackage{xcolor}

\begin{document}

\preprint{APS/123-QED}

\title{Single-electron states of phosphorus-atom arrays in silicon}

\author{Maicol A. Ochoa}
\email{maicol@nist.gov}
\affiliation{National Institute of Standards and Technology,Gaithersburg,Maryland, USA}
\affiliation{Department of Chemistry and Biochemistry, University of Maryland, College Park,Maryland, USA}

\author{Keyi Liu}
\affiliation{National Institute of Standards and Technology,Gaithersburg,Maryland, USA}
\affiliation{Joint Quantum Institute, University of Maryland, College Park, Maryland, USA}

\author{Micha\l{} Zieli{\' n}ski}
\affiliation{Institute of Physics, Faculty of Physics, Astronomy and Informatics, Nicolaus Copernicus University, ul. Grudziadzka 5, 87-100 Toru{\' n}, Poland}

\author{Garnett W. Bryant}
\email{garnett.bryant@nist.gov}
\affiliation{National Institute of Standards and Technology,Gaithersburg,Maryland, USA}
\affiliation{Joint Quantum Institute, University of Maryland, College Park, Maryland, USA}



%
%

\date{\today}

\begin{abstract}
We characterize the single-electron energies and the wavefunction structure of arrays with two, three, and four phosphorus atoms in silicon by implementing atomistic tight-binding calculations and analyzing wavefunction overlaps to identify the single-dopant states that hybridize to make the array states. The energy spectrum and wavefunction overlap variation as a function of dopant separation for these arrays shows that hybridization mostly occurs between single-dopant states of the same type, with some cross-hybridization between $A_1$ and $E$ states occurring at short separations. We also observe energy crossings between hybrid states of different types as a function of impurity separation. We then extract tunneling rates for electrons in different dopants by mapping the state energies into hopping Hamiltonians in the site representation. Significantly, we find that diagonal and nearest neighbor tunneling rates are similar in magnitude in a square array. Our analysis also accounts for the shift of the on-site energy at each phosphorus atom resulting from the nuclear potential of the other dopants. This approach constitutes a solid protocol to map the electron energies and wavefunction structure into Fermi-Hubbard Hamiltonians needed to implement and validate analog quantum simulations in these devices.
\end{abstract}

\maketitle



\section{Introduction}\label{sec:int}

Donor-based quantum devices in silicon are ideal platforms for the solid-state implementation of quantum materials and quantum simulators\cite{zwanenburg2013silicon}. After the original proposal of solid-state quantum computing in impurity-based silicon nanostructures\cite{kane1998silicon, vrijen2000electron,hollenberg2006two}, several initial attempts to fabricate these structures appeared in the literature\cite{o2001towards, pla2013high, dehollain2014single}. It was clear that one of the biggest challenges was the need for atomic precision in the fabrication of phosphorus arrays in silicon. Modern nanofabrication techniques allow for near-atomic precision in dopant placement in silicon, providing fine geometric control of the device electronic quantum states\cite{fuechsle2012single,buch2013spin,wyrick2019atom, stock2020atomic,ivie2021impact,campbell2023quantifying}. Recent reports demonstrate the experimental realization of an extended Fermi-Hubbard model in a $3 \times 3$ array of single-phosphorus quantum dots\cite{wang2022experimental}, and quantum simulations of the Su-Schrieffer-Heeger model in P-doped silicon devices\cite{kiczynski2022engineering}.

For a single donor embedded in silicon, the six-fold degeneracy of the conduction band minima splits the otherwise simple spectrum for a single, bound electron in the spherical potential of the donor. As a result, the 1$s$ electronic state splits into six states, characterized by the silicon tetrahedral symmetry as a single $A_1$, triple $T_2$, and double $E$ states. This splitting has been experimentally observed via infra-red absorption spectra\cite{picus1956absorption} and other optical measurements\cite{aggarwal1964optical}.  Valley splitting persists in the electronic structure of devices with more than one donor, leading to the existence of numerous bound states for a single electron in donor arrays\cite{klymenko2014electronic,Ochoa2}. This was already anticipated in the early studies by Luttinger and Kohn\cite{kohn1955theory, luttinger1955motion}, and later investigations on the theory of one and two donors in silicon\cite{baldereschi1970valley, saraiva2015theory}, based on multivalley effective mass theory \cite{gamble2015multivalley}.  

Understanding the charge distribution and energies of bound electron states in terms of the impurity number and geometry is critical for the implementation of quantum simulations\cite{rahman2009atomistic,le2017extended} and charge qubits\cite{hu2005charge,koiller2006electric} in dopant-based devices.  For instance, phosphorus dimers in silicon could be used to realize qubits with control based on fine tuning the charge states\cite{rahman2011stark}. Faithful analog simulations of Fermi-Hubbard models require a clear identification of site energies, electron tunnelings and on-site and long-range interactions between electrons in the dopant array. This information is encoded in the electronic structure. For the phosphorus dimer, Ref.\ \citenum{le2017extended} reports effective mass theory estimations of tunnel couplings as half of the energy separation between the symmetric and antisymmetric combinations of $A_1$ states. More recently, tunneling rates between identical and different pairs of single-impurity states were obtained from computations of intra and interorbital hopping integrals utilizing Bardeen's tunneling theory\cite{gawelczyk2022bardeen}.

In this paper, we develop a systematic approach to unveil the structure of electron states in few atom arrays that allows us to extract tunneling energies, and on-site energy corrections originating from the nearby impurities' attractive potential. We first use atomistic tight-binding theory to determine the electronic states of the multidopant arrays. For each dopant in the array, we find the electronic states bound to that dopant from tight-binding calculations. For each array state, we use a wavefunction overlap analysis to determine which site-bound single-dopant states contribute to the array states. From that, we are able to extract the tunneling and on-site energies for the single-particle part of a Hubbard model for the array. This ensures that the simple Hubbard will faithfully represent the low-energy states of the array. Our analysis reveals that the formation of hybrid states in few-atom, few-site arrays from single-impurity bound states is restricted by the wavefunction symmetry and presents energy crossings as a function of impurity-impurity separation. Moreover, this approach permits the identification of complex hybrid states where more than one type of single-impurity bound state defines the array state. Since our approach depends on the wavefunction overlap, this methodology can only provide insights on the formation of  states on arrays with separations larger than the radius of the electron state bound to single impurities. 

The organization of the paper is as follows. In Sec.\ \ref{sec:method} we describe the atomistic tight-binding calculations and the wavefunction overlap analysis. Then, in Sec.\ \ref{sec:examples} we apply this methodology to selected systems with two, three and four impurities. Finally, we summarize and conclude in Sec.\ \ref{sec:conclusion}.

\section{Wavefunction overlap analysis}\label{sec:method}
We determine the electron energies and wavefunctions in dopant arrays by implementing atomistic tight-binding calculations that reproduce the experimentally verified energy band gaps and effective masses for the relevant bands. Specifically, we implement the empirical $s p^3 d^5 s^*$ tight-binding model with spin for Si with TB parameters introduced by Boykin, Klimeck and Oyafuso in Ref.\ \cite{boykin2004valence}. Each phosphorus atom in the array replaces a Si atom introducing a confinement potential for the additional electron that we model as a screened Coulomb potential
\begin{equation}
  \label{eq:Up}
  U_P(\vec{r}) =\left\{
  \begin{array}{ll}
    -\frac{e}{4 \pi \varepsilon_{\rm Si} |\vec{r}-\vec{r}_P|} & \vec{r} \neq \vec{r}_P\\
    U_{\rm CCC} & \vec{r} = \vec{r}_P
  \end{array}
  \right. ,
\end{equation}
where $\varepsilon_{\rm Si}$ is the silicon dielectric constant, $\vec{r}_P$ is the impurity location and $U_{\rm CCC}$ is the central cell correction. For a single P atom, this model reproduces the known\cite{feher1959electron} valley-split single-electron energies with correct multiplicity and energy ordering ( $\varepsilon_{A_1} < \varepsilon_{T_2} <  \varepsilon_{E}$, with $A_1$, $T_2$ and $E$ representing the different valley-split 1s states in a single phosphorus atom)\cite{liu2023} when the central cell correction $U_{\rm CCC}$ is set to $ -3.5$ eV. We fix $U_{\rm CCC}$ to this value in our simulations of P-arrays. The dielectric constant $\varepsilon_{\rm Si}$ is set to $10.8 \varepsilon_o$. Our calculations do not incorporate the effects of strain or variations in the dielectric constant near the impurity location. The total confinement potential induced by the P-array is the sum all individual contributions Eq.\ \eqref{eq:Up}.

We analyze the single-electron energy spectrum in an array with $n$ phosphorus atoms in the following steps:
\begin{enumerate}
\item We do a tight binding calculation for the energies and wavefunctions of the array single-particle states and tight-binding calculations for the single-particle states bound to each dopant in the array.
\item For each array state, we calculate its overlap with all of the bound states of all of the single-dopants.
\item We sort array states according to their overlap with single impurity electron states. For example, we find the array states that have dominant overlap with the single-dopant states with spin-up, $A_1$ character. This allows us to identify classes for the array states. We assume that the arrays states in the same class are the group of states which can hybridize together. We expect this to be true except possibly near an anticrossing where the weaker overlaps may provide the channel for coupling.
\item We assume that the array states in the same class can be described by a single-electron Hubbard model. We identify the minimal single-electron Hubbard model in the site-representation that can provide an energy spectrum sharing the same properties as the energy spectrum of the symmetry class of array states.
\item For each class of arrays states, we find the Hubbard model parameters by fitting the eigenvalues obtained for the Hubbard model to the tight-binding energies. This gives us a way to define hopping parameters and on-site energies for each class of array states. We do this as a function of dopant-dopant separation in the array to determine the dependence of the hopping and on-site energy on separation.
\end{enumerate}

Specifically, choosing a basis set for electron states in a single P atom, $\{ | \Phi_i^\sigma \rangle \}$, we evaluate the overlap integral between each array state $| \Psi_k^{\tilde \sigma} \rangle$ and single P states $| \Phi_i^\sigma \rangle$ at each dopant position in the array. In our notation, $i$ and $k$ are wavefunctions indices and $\sigma$ and $\tilde \sigma$ indicate the electron spin in each state. We order the basis sets such that each consecutive pair is a spin-conjugate pair, i.e., we include the $i$th spin-cojugate pair $\{ |\Phi_{2i}^{\sigma_{2i}} \rangle\, , \,|\Phi_{2i+1}^{\sigma_{2i+1}}\rangle\}$ in the $2i$th and $2i+1$th positions so that  $\sigma_{2 i} = - \sigma_{2i+1}$. Before computing the overlap integral $\langle \Phi_i^{\sigma,\; (\alpha)} | \Psi_k^{\tilde \sigma} \rangle$ , we align the spins of the single impurity spin-conjugate pair $|\Phi_{2i}^\sigma \rangle$, $|\Phi_{2i+1}^{ \sigma} \rangle$ to coincide, to the best possible, with the spin orientation of the target spin-conjugate P-array states, $|\Psi_{2i}^\sigma \rangle$, $|\Psi_{2i+1}^{ \sigma} \rangle$. For each site $\alpha$ in the array, we then collect these overlaps  $M^{(\alpha)}$ with
\begin{equation}
  \label{eq:histkj}
  M_{ki}^{(\alpha)}(\vec{R})=|\langle \Phi_i^{\sigma,\; (\alpha)} | \Psi_k^{\tilde \sigma} \rangle |^2.
\end{equation}
The overlaps $M^{(\alpha)}$ depend on the array geometry, dopant position $\vec{R}$, and relative spin orientation. In this approach, we read the overlap in the form of histogram maps to separate array states into subgroups that overlap with one or two single P orbitals. For parameter calculations, we only use the P-arrays energies as detailed below.

Next, we write an $n$-dimensional single-particle Hamiltonian $\hat H_{\rm P-array}$ in the {\sl site representation} consisting of a single site energies ${\varepsilon_\alpha}$, inter-site tunneling energies $\{ t_{\alpha, \beta} \}$, and on-site energy shifts $\{ \lambda_\alpha \}$, and obtain exact forms for their eigenvalues and eigenvectors. In this case, $\alpha$ and $\beta$ are again indices listing the array impurities or, equivalently, array sites.  $\hat H_{\rm P-array}$ is therefore the {\sl site representation} of the P-array single electron states. While the tunneling rates $t_{\alpha, \beta}$ result in the formation of hybrid states between different sites, the origin of the on-site correction energies $\lambda_{\alpha}$ is in the nonuniform nature of the local potential at each site due to the impurity potential, Eq.\ \eqref{eq:Up}, from all of the sites.  On-site energy shifts $\varepsilon_\alpha \to \varepsilon_\alpha - \lambda_\alpha$, are the sum of the Coulomb potentials at the site $\alpha$ due to all impurities forming the array.   The parameter set defining the Hamiltonian model should preserve the symmetry of the array in the silicon matrix, reducing the total number of independent on-site shifts and tunneling energies in the model. Symmetry considerations also simplify the analytical forms for the $\hat H_{\rm P-array}$ eigenvalues and separate their eigenstates into subsets invariant under different array symmetry elements.

In the last step, we determine $t_{\alpha, \beta}$ and $\lambda_\alpha$ by numerically fitting the model eigenvalues to the corresponding array energies, replacing $\varepsilon_\alpha$ by the energy of the corresponding single-P orbital on each site.

When a single distance parameter $d$ characterizes the P-array, we model the functional dependence of tunneling and on-site energy shifts by the exponential forms. 
\begin{align}
  t_{\alpha,\beta}(d) =&  t_ {\alpha, \beta}^o e^{-d/l_t^{(\alpha, \beta)}} \label{eq:tx} \\
  \lambda_\alpha(d) =& \lambda_\alpha^o e^{-d/ l_\lambda^{(\alpha)}} \label{eq:lx},
\end{align}
with decay lengths $l_t^{(\alpha,\beta)}$ and $l_\lambda^{(\alpha)}$. We notice that for long lengths,  $l_{\lambda}^{(\alpha)}$,
\begin{align}
  \lambda_\alpha(d) \to& \frac{\lambda_\alpha^o}{1+d/l_{\lambda}^{(\alpha)}} \label{eq:lxlong}.
\end{align}
The form for the on-site energy correction originates from the screened potential used to represent the impurities in our model Eq. \eqref{eq:Up}. 
Occasionally, the form in Eq.\ \eqref{eq:lxlong} reproduces better the observed trends in the energies than Eq.\ \eqref{eq:lx}.

We remark that due to the Si spin-orbit coupling, the spin alignment is not always achieved with great precision. Occasionally, this results in the misalignment of the spin orientation of a single impurity spin-conjugate pair relative to the class of P-array states beyond a desired tolerance. However, the actual value of the overlap between different states is irrelevant for the calculation protocol of the system's parameters described above. Only the relative magnitude of the overlaps matters so that the classes of array states that hybridize can be identified. In our approach, we read the overlap in the form of histogram maps to separate array states into subgroups that overlap with one or two single P orbitals. To determine the Hubbard model parameters, we only use the array energies.

\section{Few-atom arrays}\label{sec:examples}
 In this section we analyze the energy spectrum for single-electron states in arrays with two, three and four phosphorus atoms. We consider the low energy array states. For this reason, we compute overlaps between the array states and the 1$s$ valley-split spin-degenerate bound states of the single dopants. For each $n$P-array we identify the lowest $12 n$ electron states with energies falling in the Si band gap and analyze their structure utilizing the methodology described in Sec.\ \ref{sec:method} 

\begin{center}
    \begin{figure}[thb]
      \includegraphics[scale=0.32]{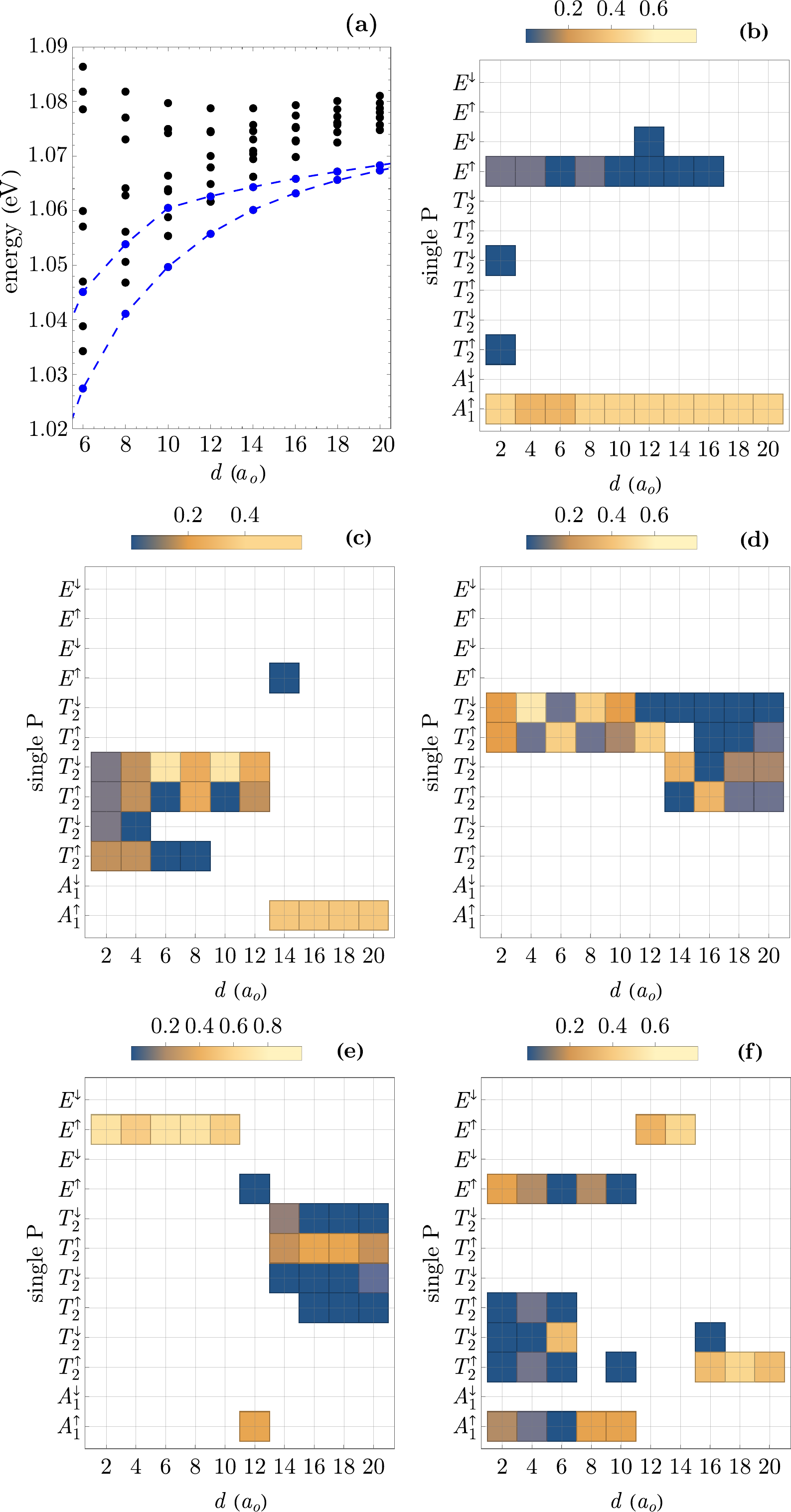}
      \caption{Phosphorus dimer overlap analysis. (a) Energy spectrum as a function of P-P separation. Each energy is doubly degenerate due to spin.  (b)-(f) Overlap histograms, Eq. \eqref{eq:histkj}, as a function of the type of single dopant state and the dimer separation $d$ for first (b), third (c), fifth (d), seventh (e), and  nineth (f) dimer state. The dimer grows along the [100] direction. Dashed lines in (a) indicate the dimer states identified with symmetry and antisymmetric combinations of single phosphorus $A_1$ electron states.  }
      \label{fig:2Pover}
    \end{figure}  
\end{center}

\subsection{Phosphorus dimer}
\begin{center}
    \begin{figure}[t]
      \includegraphics[scale=0.6]{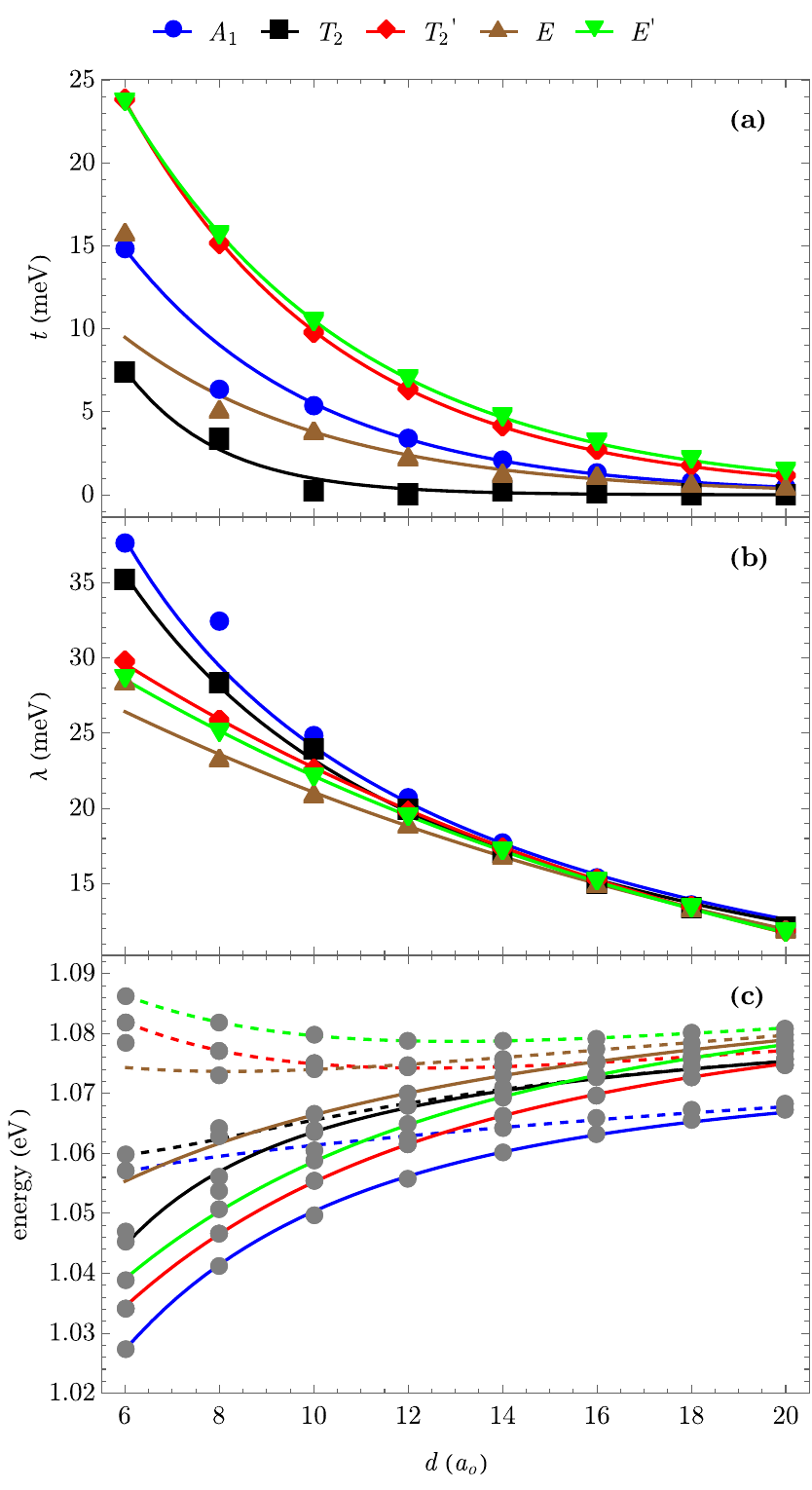}
      \caption{The phosphorus dimer along the [100] direction. (a) single-electron tunneling energy $t$, (b) on-site energy shift $\lambda$, and (c) single-electron energy spectrum as a function of P-dimer separation $d$ along the [100] direction. In Fig.\ (c), solid and dashed lines show the corresponding symmetric and antisymmetric energies $\varepsilon_{\rm 2P}^{-}$ and $\varepsilon_{\rm 2P}^{+}$ for each case of dimer-state pair identified through the overlap analysis. }
      \label{fig:2Ppar}
    \end{figure}  
\end{center}

 We begin by investigating the single electron energies for a phosphorus dimer, located along the [100] direction, in Figs.\ \ref{fig:2Pover} and \ref{fig:2Ppar}. In Fig.\ \ref{fig:2Pover}a we show how the electron energies vary as a function of the impurity separation $d$, from six to twenty $a_o$, where $a_o$ is the Si lattice constant. The energy levels for a single electron spread more as the separation decreases, with a drop in the ground state energy of several tens of meV relative to the corresponding ground state energy for an isolated phosphorus atom, which for this TB model corresponds to 1.079 eV.  We analyze the energy distribution and wavefunction structure for this dimer utilizing the overlap analysis described in Sec.\ \ref{sec:method}. Counting 2P wavefunctions from the lowest to the highest in energy, adding labels $\uparrow$, $\downarrow$ to differentiate between spin conjugate states, Figs.\ \ref{fig:2Pover}b-f show overlap histogram maps for the 1st, 3rd, 5th, 7th and 9th dimer wavefunctions. Each histogram map lists on its vertical axis twelve single-phosphorus states -- corresponding to the two-fold $A_1$ state, six-fold $T_2$ and four-fold $E$ states -- and the horizontal axis lists different P-P separations along the [100] direction. From Fig.\ \ref{fig:2Pover}b, we find that the dimer ground state overlaps mostly with the $A_1$ state at every dimer interatomic distance, with $M_{A_1 \; 1} \approx 0.5$, and small but still relevant overlap with one $E$ state for distances shorter than 8 $a_o$. In fact, $M_{A_1, 1}$ is consistently larger than $0.5$, increasing as the impurity separation decreases. Indeed, the wavefunction for an $A_1$ state centered at a single impurity spreads in space with nonneglible overlap in the neighborhood of the other impurity. As a consequence, this additional contribution to the total overlap is larger for closer impurities and it is a signature of the nonorthogonal character of the single-phosphorus wavefunctions. Even when the single-donor wavefunctions are nonorthogonal, the overlaps still provide a good way to characterize and group the dimer states. Moreover, the dimer ground state can be represented by the linear combination of $A_1$ states from each impurity in the dimer for $d \geq 10 a_o$. Considering higher energy dimer states, Figs \ref{fig:2Pover}c-f, we find that the $A_1$ state overlaps with the 3rd, 7th, and 9th states for the corresponding separations of $ d \geq 14 a_o$, $d=12 a_o $ and $d \leq 10 a_o$. We note that 5th dimer wavefunction is orthogonal to the $A_1$ state for the whole set of dimer configurations considered. Tracing out how the overlap of the single particle $A_1$ states varies with $d$, as illustrated by the blue line in Fig \ref{fig:2Pover}a, we identify the dimer states which we associate with the symmetric and antisymmetric hybridizations of single phosphorus $A_1$ states. 

We analyzed the overlap histogram maps for the lowest twenty four bound states in phosphorus dimers as we vary the impurity separation, identifying the energies associated with symmetric and antisymmetric linear combinations of $T_2$ and $E$ states. For instance, in Figs.\ \ref{fig:2Pover}e and \ref{fig:2Pover}f, we observe dimer states that overlap with one single-donoer $E$ state. Specifically, the 7th and the 9th dimer states overlap with the $E$ state at corresponding separations of $d \leq 10a_o $ and $d = 12,14 a_o$ . We also find that the 23th dimer state overlaps, for the range of dimer separations considered, with the same single-donor $E$ state (not shown in Fig. \ref{fig:2Pover}).  In this way, we identify the dimer energy states corresponding to symmetric and antisymmetric combinations of single-donor $E$ states.

The identification of symmetric and antisymmetric states permits a quantitative determination of electron tunneling rates, and the on-site energy shifts, needed for a Hubbard model, by a fitting to atomistic tight-binding energy calculations. The Hamiltonian for the two-site representation of the dimer, with on-site energy $\varepsilon_{\rm P} \in \{ \varepsilon_{A_1},\varepsilon_{T_2},\varepsilon_{E}\} $, tunneling energy $t$, and on-site shift energy $\lambda$ is 

\begin{equation}
  \label{eq:H2P}
  \hat H_{\rm 2P} =
  \begin{pmatrix}
    \varepsilon_{\rm P} - \lambda & - t \\
    -t &
\varepsilon_{\rm P} -\lambda
   \end{pmatrix}\; ,
\end{equation}
and has eigenvalues $\varepsilon_{\rm 2P}^{\pm} = \varepsilon_{\rm P} - \lambda \pm t $, with corresponding eigenstates $v_{\rm 2P}^{\mp}=(1/\sqrt{2},\mp 1/\sqrt{2})$. In this form, the symmetric state $v_{\rm 2P}^{+}$ corresponds to the lowest energy state $\varepsilon_{\rm 2P}^-$. By identifying $\varepsilon_{\rm 2P}^{\pm}$ with the symmetric and antisymmetric energies obtained from the wavefunction overlap analysis, we compute  $t$, and  $\lambda$. 

\begin{center}
    \begin{figure}[t]
      \includegraphics[scale=0.6]{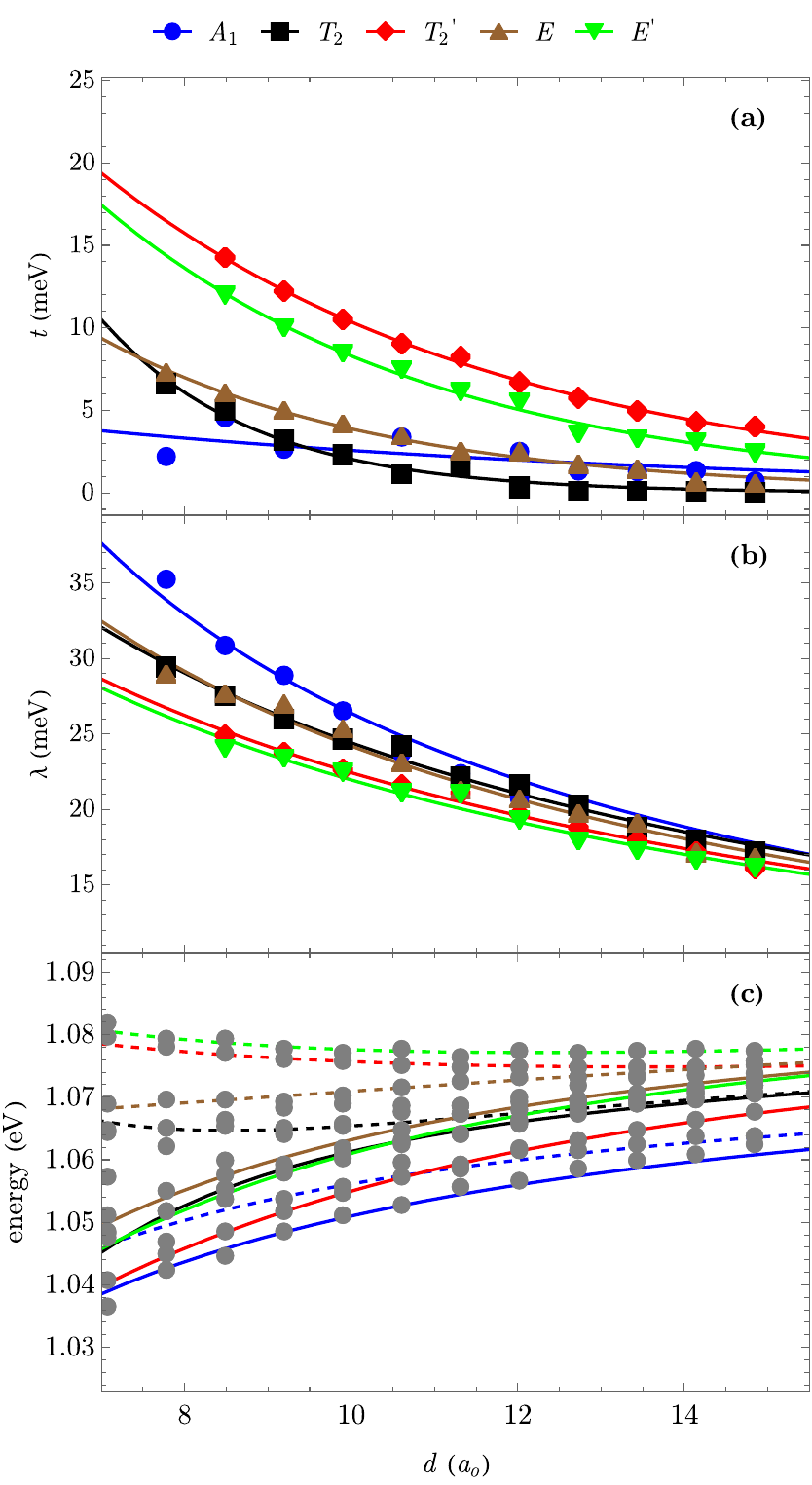}
      \caption{The phosphorus dimer along the [110] direction. (a) single-electron tunneling, (b) on-site energy shift, and (c) single-electron energy spectrum as a function of P-dimer separation $d$ along the [100] direction.  In Fig.\ (c), solid and dashed lines show the corresponding symmetric and antisymmetric energies $\varepsilon_{\rm 2P}^{-}$ and $\varepsilon_{\rm 2P}^{+}$ for each case of dimer state pair identified through the overlap analysis.}
      \label{fig:2P110par}
    \end{figure}  
\end{center}

We show the result for these two parameters in  Figs.\ \ref{fig:2Ppar}a and  \ref{fig:2Ppar}b, where we also include fitting curves corresponding to the expressions in Eqs.\ \eqref{eq:tx} and \eqref{eq:lx}. Significantly, we observe in Fig.\ \ref{fig:2Ppar}a that the assumed exponential form for the tunneling rate coincides with the $t$-values obtained from the atomistic tight-binding calculations, only showing deviations at separations less or equal to $8 a_o$ for the $A_1$ and $E$ states. Indeed, for these separations, both $A_1$ and a single $E$ states overlap with the same set of dimer states (see, Fig.\ \ref{fig:2Pover}(b-f), suggesting that in this case the form of the dimer state falls beyond the symmetric/antisymmetric representation and should include linear combinations of both $A_1$ and $E$ single-donor states. Our results in Fig. \ref{fig:2Pover}a also reveal that the tunneling rates for the ground state, the symmetric combination of $A_1$ states, is significantly larger than previously reported\cite{gamble2015multivalley}, whenever the tunnel coupling has been estimated as the difference between the first excited and ground states\cite{hu2005charge}. This better estimation  of $t$ results from the correct identification of energy crossings at shorter P-P separations, which are fully identified by our wavefunction overlap analysis. Figure \ref{fig:2Ppar} demonstrates that the on-site shifts are significant for the $d$ range analyzed. The shifts are larger in magnitude than the tunneling energies at each separation, and that they follow the exponential form in Eq.\ \eqref{eq:lx}. Finally, introducing the $t$ and $\lambda$ fitting forms in $\varepsilon_{\rm 2P}^{\pm}$, we obtain the functional dependence on the impurity-impurity distance for the electron energies. Figure \ref{fig:2Ppar}c shows that for the donor separations investigated, this results in an excellent agreement between the tight-binding energies and a site model for the dimer with eigenstates corresponding to symmetric and antisymmetric combinations of single P-states. Consequently, we quantify the geometric modulation of tunneling energies in a phosphorus dimer in a form that makes these parameters useful for Hubbard models.

Next, we carry out the overlap analysis on P dimers along the [110] direction. The result is presented in Fig.\ \ref{fig:2P110par}. We find that tunneling rates for the $T_2$ and $E$ states anticipated by the formation of symmetric and antisymmetric states follow the exponential form in Eq.\ \eqref{eq:tx} for $8 a_o < d < 15 a_o$, in contrast with the tunneling rate between $A_1$ states, since the latter display oscillations as a function of the dimer separation (Note in Fig.\ \ref{fig:2P110par}c how the ground tight-binding state energy, represented by solid dots, oscillates around the solid blue line). Similarly to the case of P-dimers distributed along the [100] direction, the $E$ and $T_2$ states separate in two kinds, with strong and weak tunneling rates. Moreover, the weakly hybridized $T_2$ dimer state is two-fold degenerate in both cases for the distance range considered. The on-site shift energies in Fig.\ \ref{fig:2P110par}b are like those found in Fig.\ \ref{fig:2Ppar}b, predicting larger shifts for the $A_1$, $T_2$ and $E$ states, in that order. However, the on-site shifts in the [110] direction were better reproduced by the rational form for $\lambda$ in Eq.\ \eqref{eq:lxlong}. Utilizing the functional forms for $t$ and $\lambda$ we obtain the $d$-dependence of dimer energies $\varepsilon_{2P}^{\pm}$ that we compared with the result of the atomistic tight-binding calculations in Fig.\ \ref{fig:2P110par}c. The agreement between the two-level model and the atomistic calculations is acceptable for $d \geq 8 a_o$. Additional improvements to the fits may require linear combinations of additional single-donor states and considering tunneling rates between different types of single-donor states (see, e.g., Ref.\ \citenum{gawelczyk2022bardeen}).

In comparison with previous results reported in the literature, we find that, for example Ref.\ \citenum{tankasala2022shallow}, provides a partial identification of symmetric and antisymmetric states based on wavefunction symmetry, for a dimer grown in the [110] direction (see Fig.\ 2 in the cited Reference). Specifically, the authors of Ref.\ \citenum{tankasala2022shallow} identify the lowest three states  as the symmetric combinations of $A_1$, $T_2$ and the antisymmetric $A_1$ state in the region between 4 and 8 nm, observing energy crossing around 8 nm separation between the symmetric $T_2$ state and the antisymmetric $A_1$ state. Our overlap analysis coincides with this observation in the range of dimer separations considered but predicts the energy crossing occurring at $\sim 5.7$ nm  ($\sim 10.5 a_o$). This difference could result from the distance of the donor pair to the silicon surface, variations in the tight-binding parameter set or central cell correction. Moreover, our overlap analysis provides a full description of the higher energy states in terms of symmetric and antisymmetric states.

\subsection{Linear phosphorus trimer}
\begin{center}
    \begin{figure}[thb]
      \includegraphics[scale=0.31]{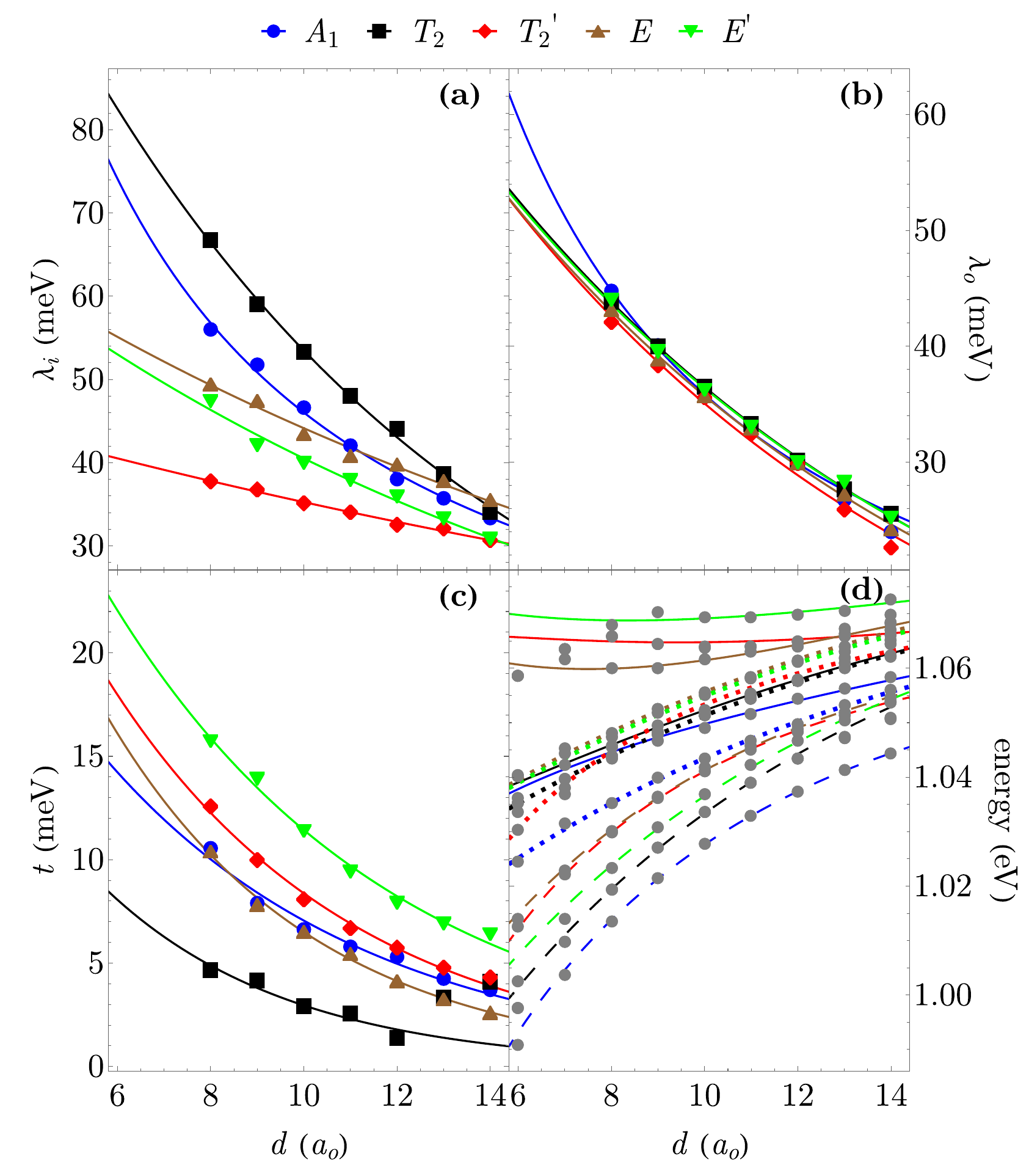}
      \caption{The linear phosphorus trimer along the [100] direction. The $d$ dependence for (a) On-site energy shift at the inner impurity, (b) on-site energy shift at the outer impurity, (c) nearest neighbor electron tunneling. (d) Single-electron energy spectrum as a function of the separation $d$ of each outer impurity from the inner one. For each case, we used solid, dotted, and dashed lines to show fits corresponding to the energies $\varepsilon^{+1}_{\rm 3P}$, $\varepsilon^{0}_{\rm 3P}$, and $\varepsilon^{-1}_{\rm 3P}$.}
      \label{fig:3P100}
    \end{figure}  
\end{center}

We now investigate a system with three phosphorus atoms, forming a line along the [100] direction where the outer impurities are at a distance $d$ from the inner atom. For this family of 3P-arrays, we perform the full overlap analysis on the lowest thirty-six bound states found from a full tight-binding simulation, with energies in the Si band gap. Specifically, we consider arrays with $d$ varying from 6 to 14 $a_o$, and report the result of the atomistic tight-binding calculation and overlap analysis in Fig.\ \ref{fig:3P100}.

For the linear 3P-array, we adopt a site-representation including different on-site corrections for inner and outer atoms, respectively $\lambda_i$ and $\lambda_o$, and a single nearest-neighbor tunneling rate $t$. The corresponding Hamiltonian 

\begin{equation}
  \label{eq:H3P}
  \hat H_{\rm 3P} =
  \begin{pmatrix}
    \varepsilon_{\rm P} - \lambda_o & - t & 0 \\
    -t & \varepsilon_{\rm P} -\lambda_i & -t\\
    0 & -t & \varepsilon_{\rm P} - \lambda_o \\
   \end{pmatrix},
\end{equation}
has three distinct eigenvalues $\varepsilon_{\rm 3P}^m$, $m \in \{-1,0,1\}$, given by
\begin{align}
  \varepsilon_{\rm 3P}^0 =& \varepsilon_{\rm P}- \lambda_o\\
  \varepsilon_{\rm 3P}^{\pm 1} =& \varepsilon_{\rm P} -\frac{\lambda_i+\lambda_o}{2} \pm \sqrt{2\, t^2 +\left( \frac{\lambda_i-\lambda_o}{2}\right)^2}.
\end{align}
The state with energy $\varepsilon_{\rm 3P}^0$ has zero amplitude at the inner impurity, and corresponds to the antisymmetric combination of the outer P states (see Ref.\ \cite{Ochoa2}). On the other hand, the symmetric combination of the outer P states hybridizes with the inner P orbital resulting in states with energies $\varepsilon_{\rm 3P}^{\pm 1}$. In addition, $\varepsilon_{\rm 3P}^{-1} \le \varepsilon_{\rm 3P}^{0} \leq \varepsilon_{\rm 3P}^{+1} $ whenever $\lambda_i \ge \lambda_o \geq 0$. Since the on-site energy shifts originate from the impurity model, that is, the screened Coulomb potential in Eq.\ \eqref{eq:Up}, we anticipate a larger on-site energy drop on the inner impurity. Only states with energies $\varepsilon_{\rm 3P}^{\pm1}$ effectively couple the outer impurities with the central one. Moreover, whenever  $\lambda_o  < \lambda_i$, the state with energy $\varepsilon_{\rm 3P}^{-1}$ will find most of its electron density localized near the inner impurity, while for the state with energy $\varepsilon_{\rm 3P}^{+1}$  the electron density will be found at the outer impurities.

For each of the $A_1$, $T_2$ and $E$ single-P states, we identified from the thirty-six overlap histogram maps the triples of 3P-array states corresponding to combinations of each single-impurity state. For instance, for $d=10 a_o$, we noted that the $A_1$ state for the inner impurity and the $A_1$ state at each outer impurity overlapped with the 1st, 13th and 15th P-trimer bound states. We ascribe the energies corresponding to these states as the energies $\varepsilon_{\rm 3P}^{-1}$, $\varepsilon_{\rm 3P}^{0}$ and $\varepsilon_{\rm 3P}^{+1}$. Indeed, we find that the first trimer state is mostly localized at the inner impurity; 13th trimer state has vanishing overlap with the inner $A_1$ state, and the 15th trimer state overlaps with local $A_1$ states at each impurity.    

Figures \ref{fig:3P100}a and \ref{fig:3P100}b show the inner and outer on-site shifts for the trimer as a function of the separation $d$. We note that among all the single P states, the outer shift $\lambda_o$ is very similar, while the inner shift $\lambda_i$ depends strongly on each state. In all cases, we observe that $\lambda_i > \lambda_o$, with the exception of the $T_2'$ hybrid states for separations $d \leq 9 a_o$. Figure \ref{fig:3P100}c presents the tunneling strength $t$, showing a good agreement between values obtained from the overlap analysis of our atomistic tight-binding calculations and the exponential form Eq.\ \eqref{eq:tx},  with the exception of the weakly hybridized $T_2$ states at separations 13 and 14 $a_o$. Significantly, when the fitted forms are substituted into the site-representation energies $\varepsilon_{\rm 3P}^m$, we find in Fig.\ \ref{fig:3P100}d a good description of the single-electron energies as a function of $d$. 

\subsection{Phosphorus square lattice}

\begin{center}
    \begin{figure}[thb]
      \includegraphics[scale=0.31]{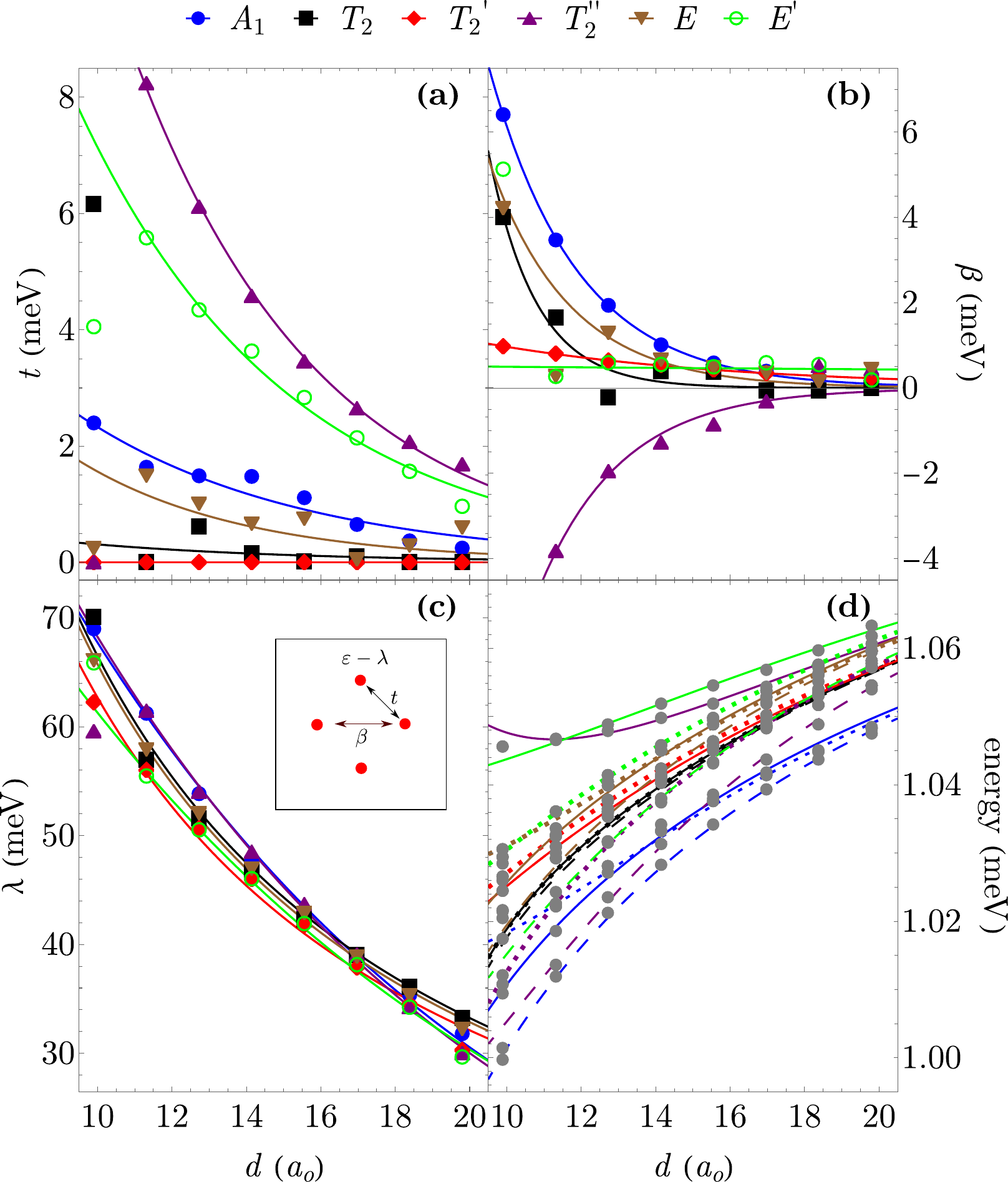}
      \caption{ The square array. (a) Nearest-neighbor tunneling energy, (b) diagonal or next-nearest-neighbor tunneling energy, (c) On-site energy shift, and (d) single-electron energy spectrum as a function of the square side length $d$.
For each case, we used solid, dotted, and dashed lines to show fits corresponding to the energies $\varepsilon^{+1}_{\rm 2P \times 2P}$, $\varepsilon^{0}_{\rm 2P \times 2P}$, and $\varepsilon^{-1}_{\rm 2P \times 2P}$. }
      \label{fig:2x2Diam}
    \end{figure}  
\end{center}

\begin{center}
    \begin{figure}[thb]
      \includegraphics[scale=0.85]{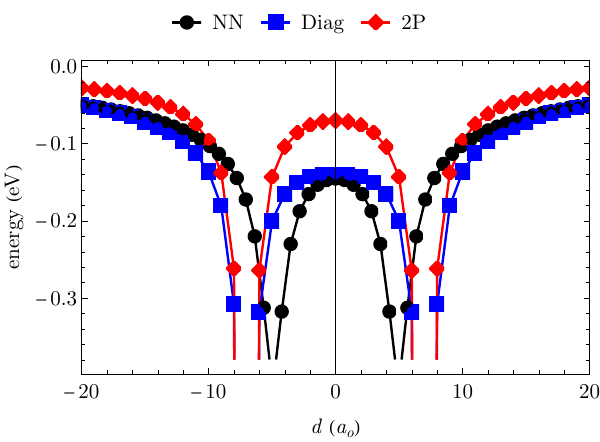}
      \caption{Confinement potential sections for an electron in a 2P$\times$2P and a 2P array. We show the potential along the line joining nearest neighbor (NN) atoms (black, circles), and the diagonal (blue, squares) in the 2P$\times$2P array with $d=7\sqrt{2}a_o$. For the 2P array formed by the P-atoms in the diagonal, we present the potential energy along the straight line (red,diamonds). }
      \label{fig:pot}
    \end{figure}  
\end{center}

In this section we study square arrays formed by four phosphorous atoms, oriented such that the diagonals are parallel to the [100] and [010] directions (see inset Fig.\ \ref{fig:2x2Diam}).  We consider a single-electron Hamiltonian in the site representation

\begin{equation}
  \label{eq:H2x2P}
  \hat H_{\rm 2P \times 2P } =
  \begin{pmatrix}
    \varepsilon_{\rm P} - \lambda &-t& -\beta& -t \\
    -t & \varepsilon_{\rm P} -\lambda & -t& - \beta\\
    -\beta & -t & \varepsilon_{\rm P} - \lambda & -t \\
    -t & -\beta& -t & \varepsilon_{\rm P} - \lambda  \\
   \end{pmatrix} ,
 \end{equation}
 accounting for electron tunneling to nearest and next-nearest neighbor impurities with corresponding tunneling energies $t$ and $\beta$. Since the impurities forming the square are equivalent in this configuration, a single on-site shift energy $\lambda$ is considered.  This Hamiltonian has three distinct eigenenergies $\varepsilon_{\rm 2P\times 2P}^{m}$, with $m \in \{-1,0,1\}$,
 \begin{align}
   \varepsilon_{\rm 2P\times2P}^{0}=& \varepsilon_{\rm P}- \lambda + \beta ,\\
   \varepsilon_{\rm 2P\times2P}^{\pm 1}=& \varepsilon_{\rm P}- \lambda \pm 2 t - \beta ,
 \end{align}
 with a two-fold degenerate energy state $\varepsilon_{\rm 2P\times2P}^{0}$. The eigenvectors for the Hamiltonian in Eq.\ \eqref{eq:H2x2P} are $v_{\rm 2P\times 2P}^{-1} = (1/2,1/2,1/2,1/2)$, $v_{\rm 2P\times 2P}^{+1} = (1/2,-1/2,1/2,-1/2)$ and, for the two-dimensional eigenspace corresponding to $\varepsilon_{\rm 2P \times 2P}^0$ we can take as a basis $v_{\rm 2P\times 2P}^{0 a} = (1/\sqrt{2},0,-1/\sqrt{2},0)$ and $v_{\rm 2P\times 2P}^{0 b} = (0,1/\sqrt{2},0,-1/\sqrt{2})$.     For this family of 4P-arrays, we choose as a geometric control parameter the square side length $d$ -- also corresponding to the nearest neighbor separation between impurities -- and carry out the overlap analysis for the lowest forty-eight single-particle array states. In this case only overlaps between a single impurity and the square-array states are evaluated because the dopants forming the square are equivalent.   Figures \ref{fig:2x2Diam}a and \ref{fig:2x2Diam}b shows the variation in $t$ and $\beta$, as a function of $d$. Clearly, both $t$ and $\beta$ are comparable in magnitude for the range of separations considered. For the square-array ground state, the nearest-neighbor tunneling is larger than the diagonal tunneling for $d \geq 14 a_o$. However, for $d < 14 a_o$, the opposite trend occurs, i.e., $\beta > t$. we anticipate enhanced  electron tunneling across the square diagonal resulting from the gating action of the transverse impurity pair, lowering the potential barrier across the diagonal. Figure \ref{fig:pot} illustrates this finding, by showing the potential across the line joining the corresponding P atoms for a separation $d = 7\sqrt{2}a_o$. The barrier to tunneling across the diagonal is significantly lower than the barrier between two P in a dimer with the same separation as the diagonal length. This is the gating effect. In fact, the barrier height for diagonal tunneling is the same, in the figure, as the barrier height for nearest neighbor tunneling. In this case, other factors, such as the effective mass along the tunneling direction determine which tunneling is more effective.

 Figure \ref{fig:2x2Diam}a and \ref{fig:2x2Diam}b also reveal that $T_2$ states form three types of hybrid states in a square lattice. The first type, labeled $T_2$ in Fig.\  \ref{fig:2x2Diam}, displays weak nearest-neighbor and next-nearest-neighbor tunneling; the second type, $T_2'$, has almost vanishing nearest-neighbor tunneling and moderate next-nearest-neighbor tunneling. In contrast, the third type $T_2''$, shows relatively significant nearest neighbor and next-nearest-neighbor tunneling energies, between one and ten meV, with $\beta$ negative. Electron states resulting from hybridizations of the single-phosphorus $E$ states also show significant diagonal tunneling rates for the square sizes considered. Moreover, Fig. \ref{fig:2x2Diam}c reveals that the variation in the on-site shifts as a function of $d$ is very similar among all the array states, with optimal exponential fits, as in Eq.\ \eqref{eq:lx}, for the $A$, $T_2''$ and $E$ states; while other cases follow the form in Eq.\ \eqref{eq:lxlong}.

 For the square array we are also able to match the tight-binding spectrum in Fig.\ \ref{fig:2x2Diam}d in the range $10 a_o \le d \le 20 a_o $ by combining the fitted forms for $t$, $\beta$ and $\lambda$ with the energies $\varepsilon_{\rm 2P\times2P}^{m}$. We conclude that in this range, the ground state for a single electron is well approximated by the linear combination of $A$ states, as $v_{\rm 2P\times 2P}^{-1}$. However, our Hamiltonian model fails to reproduce the high energy states in the spectrum when $d < 12a_o$.\\

\section{Summary}\label{sec:conclusion}

We introduced a systematic approach for calculating electron tunnelings and on-site energy shifts for single-electron states in P-doped devices from atomistic tight-binding wavefunctions. As a consequence of valley splitting, P-arrays display numerous single-electron bound states. Starting from the wavefunction overlap between the array and single-impurity states, we described each electron wavefunction as a linear combination of just a few single-phosphorus electron states localized at each impurity. Interpreting the overlap maps for the relevant portion of the spectrum utilizing model site Hamiltonians, we extracted tunnelings and on-site energy shifts as a function of the geometric parameters, matching the energy spectrum of a full, atomistic tight-binding calculation. This approach is relevant for interpreting analog quantum simulations of the Fermi-Hubbard model in Si:P devices. Remarkably, we demonstrated that nearest-neighbor and next-nearest-neighbor tunneling energies can be of the same magnitude in a square lattice.

The overlap analysis introduced here cannot resolve the structure of electron states in clusters -- arrays where impurity-impurity separation is similar or smaller than the wavefunction radius-- and can only provides qualitative information on this strongly hybridized regime. In this regard, we remark that our approach to obtain model parameters from the single-electron energies does not depend on the actual values of the overlaps. The overlaps are only used as a indicator of which single-impurity states contribute to an array state. In turn, this tells us which array states are derived from the same type of impurity states and this tells us which states are coupled together by the same hopping Hamiltonian. We directly find effective hopping parameters for a Hamiltonian that reproduces the full atomistic calculations. Our approach using the overlap analysis can be extended to calculate electron-electron interactions if we first obtain the many-electron states of the system and do the overlap analysis on those states.

\begin{acknowledgments}
M.Z. acknowledges support from the Polish National Science Centre based on decision No. 2015/18/E/ST3/00583
\end{acknowledgments}


\bibliography{Parray.bib}

\end{document}